\documentclass[review,5p]{elsarticle}
\usepackage[utf8x]{inputenc}
\usepackage{lineno}
\usepackage{amssymb,amsmath}
\usepackage[usenames]{color}
\usepackage{hyperref}
\usepackage{multirow}

\journal{Astronomy and Computing}

\biboptions{authoryear}

\begin{document}

\begin{frontmatter}

\title{Two-Fluid Dusty Gas in Smoothed Particle Hydrodynamics: Fast and Implicit Algorithm for Stiff Linear Drag}

\author[mymainaddress,ivt,nsu]{Olga P. Stoyanovskaya\corref{mycorrespondingauthor}}
\ead{stop@catalysis.ru}

\author[mymainaddress,nsu]{Tatiana A. Glushko}

\author[mysecondaryaddress]{Nikolay V. Snytnikov}

\author[mymainaddress,nsu]{Valeriy N. Snytnikov}
\cortext[mycorrespondingauthor]{Corresponding author}

\address[mymainaddress]{Boreskov Institute of Catalysis, Lavrentieva, 5, 630090, Novosibirsk, Russia, }
\address[ivt]{Institute of Computational Technologies, Lavrentieva, 6, Novosibirsk, 630090, Russia}
\address[nsu]{Novosibirsk State University, Pirogova, 2 Novosibirsk, 630090, Russia}
\address[mysecondaryaddress]{Institute of Computational Mathematics and Mathematical Geophysics, Lavrentieva, 6, Novosibirsk, 630090, Russia}

\begin{abstract}

Simulation of the dynamics of dust-gas circumstellar discs is crucial in understanding the mechanisms of planet formation. The dynamics of small grains in the disc is stiffly coupled to the gas, while the dynamics of grown solids is decoupled. Moreover, in some parts of the disc the concentration of the dust is low (dust to gas mass ratio is about 0.01), while in other parts it can be much higher. These factors place high requirements on the numerical methods for disc simulations. In particular, when gas and dust are simulated with two different fluids, explicit methods require very small timestep (must be less than dust stopping time $t_{\rm stop}$ during which the velocity of a solid particle is equalized with respect to the gas velocity) to obtain solution, while some implicit methods requires high temporal resolution to obtain acceptable accuracy. Moreover, recent studies underlined that for Smoothed particle hydrodynamics (SPH) when the gas and the dust are simulated with different sets of particles only high spatial resolution $h<c_{\rm s} t_{\rm stop}$ guaranties suppression of numerical overdissipation due to gas and dust interaction.

To address these problems, we developed a fast algorithm based on the ideas of (1) implicit integration of linear (Epstein) drag and (2) exact conservation of local linear momentum. We derived formulas for monodisperse dust-gas in two-fluid SPH and tested the new method on problems with known analytical solutions. We found that our method is a promising alternative for the previously developed two-fluid SPH scheme in case of stiff linear drag thanks to the fact that spatial resolution condition $h<c_{\rm s} t_{\rm stop}$ is not required anymore for accurate results.

\end{abstract}


\end{frontmatter}

\section{Introduction} \label{sec:intro}
Simulation of the dynamics of circumstellar discs is crucial in understanding the mechanisms of planet formation. The state-of-the-art models of formation and evolution of circumstellar discs include a variety of physical processes: dynamics of a two-phase medium (gas and solid particles) in the field of the star and in their own gravitational field, radiation transfer, chemical processes, effects of magnetic hydrodynamics and other phenomena. The review by \citet{HaworthEtAl2016} describes  applied approaches and newest computational challenges for this field. For the simulation of disc dynamics such software solutions as GANDALF \citep{GANDALF},  PHANTOM \citep{Phantom2017}, GADGET \citep{Gadget2}, ZEUS \citep{StoneNorman1992}, FEoSaD \citep{VorobyovBasu2006}, Sombrero \citep{1MNRAS,SPHinCEll} and many others are developed. They are run on supercomputers with shared and distributed memory. The selection of parallelization strategies is determined by the supercomputer hardware architecture and methods of solution of the model core equations. Additional physical and chemical processes can be included in the core model and developed parallelized codes by the operator splitting technique with use of the algorithms, requiring only minor changes in the core code. 

An example of this approach is the inclusion of the dust dynamics in the supercomputer codes for simulation of formation and evolution of the circumstellar discs. At the early evolution stages the circumstellar disc matter consists above 95\% gas and one up to several \% dust (e.g. \cite{WilliamsBest2014}). Gas dynamics processes in discs (for example, the formation of gas giants, episodic bursts of young stars) are modeled on the base of Navier-Stokes equations with turbulent viscosity. However, computer simulations of terrestrial planet formation require the introduction of dust dynamics models together with gas dynamics. Tiny dust particles and centimeter size objects are often simulated in the disc model with Euler equations for gas dynamics with zero pressure. Gas and dust are dynamically coupled by gravitation field and interact through drag force. Drag force causes the momentum transfer between dust and gas. The computing of drag force between gas and dust in the circumstellar discs appeared to be a sophisticated computational challenge. Its highlights are given in Section \ref{sec:survey}. 

In this paper we introduce a novel method for treating gas and dust mixture in two-fluid Smoothed particle hydrodynamics simulation (TFSPH).

This method suggests that gas and dust medium is simulated with two different sets of particles. 
Contrary to previous TFSPH methods (e.g. \cite{MonaghanKocharyan1995,BateDust2014}), the scheme both conserves momentum locally and allows computationally fast semi-implicit integration. The latter grants overcoming prohibitively strict temporal and spatial resolution criteria known for small grains (see Section \ref{sec:survey} for details). Moreover, the method allows for inclusion of the dust dynamics into parallelized codes for gas discs simulation on the base of SPH. 

In Section \ref{sec:model} we specify the continuity equations and equations of motions of the gas and in a circumstellar disc. In Section \ref{sec:DragINTFSPH} we describe the known approaches to the computing of drag force for the two-fluid SPH and outline our own developed method. In Section \ref{sec:methods} we specify the tested numerical schemes for the problem. In Sections \ref{sec:DustyWave} and \ref{sec:DustyShock} we describe the test problems, and in Section \ref{sec:resultsDiscussion} we present the results of calculations. Conclusions and directions for further research are summarized in Section \ref{sec:resume}. 

\section{Numerical simulation of dust dynamics in circumstellar discs. Existing approaches and problems} 
\label{sec:survey}

The dust stopping time $t_{\rm stop}$, during which the velocity of a solid particle is equalized with respect to the gas velocity, presents a characteristic time scale of the problem. If explicit methods are used for the integration of the equations of motion for a separate particle or for a dust cloud, then the time step $\tau$ must satisfy the condition (see, e.g. \citet{StoyanovskayaDust}) 
\begin{equation} 
\label{eq:timeresSPH} 
\tau < 2t_{\rm stop}. 
\end{equation} 
The dust stopping time $t_{\rm stop}$ depends on the size of a particle. In the Epstein drag mode of the diluted gas flow about a solid body (according to \citet{Epstein1924} $s<2.25\lambda$, where $\lambda$ is the mean free path of a gas molecule, $s$ is the radius of a spherical dust particle) it is determined by the equation (e.g.\citet{RiceEtAl2004}):
\begin{equation} 
\label{eq:t_stop_rho} 
t_{\rm stop}=\frac{s \rho_{\rm s}}{c_{\rm s} \rho_{\rm g}},
\end{equation} 
where $\rho_{\rm s}$ is the material density of the dust (not a volume density), $\rho_{\rm g}$ is the volume gas density in the circumstellar disc, $c_{\rm s}$ is the sound speed in a gas. 

At early stages, the dust particles in the disc have typical sizes from 1 $\mu$m to few cm. For the particles with sizes about 1 $\mu$m the time $t_{\rm stop}$ in the disc is about 100 sec (see, for example, \citet{StoyanovskayaDust,LaibePrice2011Test}); however, we want to simulate the disc dynamics for $10^4$~years or more. This means, that in view of (\ref{eq:timeresSPH}), the numerical solution of the nonstationary equations for the gas and dust medium, expanded for typical times of the disc dynamics with explicit integration schemes would require about $10^{10}$ time steps. To get rid of such a huge computational costs a quasistationary approach is often used for the simulation of tiny dust particles. In this approach, the dust velocity at the moment $\tau \gg t_{\rm stop}$ is calculated from the algebraic quasistationarity condition, relating the dust and gas velocities, rather than from the solution of a differential equation of motion. On the other hand, for bodies with sizes about 1~cm $t_{\rm stop}$ is of the order of $10^6$~sec, which is compatible with the orbital period of a particle with orbital radius of 1~$au$. For those bodies the quasistationary approach is unjustified. 

Therefore, the need for development of numerical methods arises, allowing for integration of the equations of motion for the dust with time step $\tau$, determined only by the Courant condition for gas dynamics, rather than by $t_{\rm stop}$. Such schemes employ implicit algorithms for drag computing, e.g. \citet{Monaghan1997}, or analytic integration of equations for the dust velocity field in every point of the space, e.g. \citet{BateDust2015}. The methods for solution of the equations of motion for the dust, based on these concepts, are used in \citet{BaiStone2010ApJS,ZhuDust,ChaNayakshinDust2011,FranceDustCode,Pignatale2016} and other works. They are systematically reviewed in \citet{StoyanovskayaDust}. 

The next challenge is the transition from dilute to dense particulate flows. If the dust to gas mass ratio $\varepsilon=\displaystyle \frac{\rho_{\rm d}}{\rho_{\rm g}}$ does not exceed several percents, it is possible to take into account only the effect of drag force from gas on dust, neglecting the backreaction of the dust with a tiny total mass on the dynamics of the massive gas. The dust to gas mass ratio $\varepsilon\approx0.01$ is characteristic for molecular clouds forming discs. However, in certain areas of the disc (spiral arms \citep{RiceEtAl2004}, inner parts of the disc \citep{VorobyovEtAl2017}, self-gravitating gaseous clumps \citep{ChaNayakshinDust2011,NayakshinDust2017}) the dust concentration with respect to gas can increase essentially. In this case, the equation for gas motion must be completed with a term, describing the exchange of momentum with dust particles. In a medium with high drag coefficient between gas and dust at elevated dust concentration an adequate computing of the momentum transfer from dust to gas is a challenge (e.g. \citet{BateDust2014,LaibePrice2014OneFluidDust,Ishiki2017,YangJohansen2016,VorobyovEtAl2017}). 

In particular, \citet{VorobyovEtAl2017} applied a grid-based method \citep{StoneNorman1992} for simulating the dynamics of gaseous and dusty components of a disc. Gas and dust characteristics were determined in the same space points. A first order implicit scheme (e.g. \cite{ChaNayakshinDust2011}) with operator splitting was applied for calculation of mutual drag between gas and dust. The numerical methods were tested on the problem of sound wave propagation \citep{LaibePrice2011Test} and on the shock tube problem \citep{Sod1978,LaibePrice2011}. It was found that for a system with mutual momentum transfer between gas and dust, the applied scheme required a time step $\tau$ much smaller than the $t_{\rm stop}$. However, if only the impact of gas on the dynamics of dust is taken into account (neglecting the backreaction of dust on the gas dynamics) the elaborated scheme yielded satisfactory results on these test problems without strict limitations for the time step used. In \citet{StoyanovskayaDustMultigrain,Ishiki2017} the examples of schemes are given, which solve this problem for grid methods in a medium consisting of a mixture of gas and monodisperse dust. 

Moreover, \citet{LaibePrice2011} have found, that for the same test problems, treated with SPH method \citep{MonaghanKocharyan1995}, obtaining solution with an acceptable level of dissipation requires too small spatial resolution. More specifically, in the numerical model \citep{LaibePrice2011} gas and dust were described by separate groups of model particles (TFSPH). If the drag coefficient between gas and dust is high enough, and the dust concentration is elevated (that is $\varepsilon\approx1$), an adequate accuracy is obtained only with the smoothing length $h$, satisfying the condition 
\begin{equation} 
\label{eq:sparesSPH} 
h<c_{\rm s} t_{\rm stop}. 
\end{equation} 

From (\ref{eq:sparesSPH}) it follows that the smoothing length for the simulation of micron size dust must be of the order of $10$~km, while a typical size of the circumstellar disc is $1.5 \times 10^{10}$~km. It is clear that such a spatial resolution is nowadays beyond the capabilities of computation hardware. 

The solution for the mentioned problems for numerical models, based on smoothed particles hydrodynamics, was suggested in \citet{LaibePrice2014OneFluidDust} with use of the transition towards a one-fluid model of the two-phase medium. In this approach, the system of equation for the two-phase gas and dust medium is formulated in the following variables: density of the carrier gas, dust to gas mass ratio, barycentric velocity of the medium and relative velocity of gas and solid particles. Therefore, instead of model particles of two types (gas and dust particles), model particles of a gas-dust medium are considered which carry the properties of both phases. 

On the other hand, the two-fluid approach has an advantage over single-fluid systems, if the drag force between gas and dust are weak, so dust particle velocities can essentially differ locally. In \citet{BateDust2014,BateDust2015,ClarkeDust2015} the authors developed an approximation within the two-fluid approach for consideration of drag force (see Section \ref{sec:DragINTFSPH} for details) and applied a semi-analytical technique for the solution of the equation of motion for dust and gas. As a result, the restrictions for selection of the smoothing length (\ref{eq:sparesSPH}) could be weakened for the simulation of media with high dust concentration of small grains. \citet{BateDust2014} have found, that the obtained dissipation in their solutions for a gas-dust medium is smaller compared to the scheme of \citet{MonaghanKocharyan1995}. However, their dissipation exceeds the dissipation in a gas due to introduction of artificial viscosity. Therefore, further development of fast and exact methods for numerical simulation of gas and dust media on the base of two-fluid smoothed particle hydrodynamics still remains a topical challenge. 

We note also that for a gas-dust medium with polydisperse dust a method for dynamical simulation MULTIGRAIN \citep{Multigrain} was developed on the base of one-fluid approach in SPH taking into account backreaction of dust on the gas velocity. Besides that, fast method to compute polydisperse dust and gas velocities was obtained in \citet{StoyanovskayaDustMultigrain} for grid-based approach. 

We underline that an important requirement for all suggested algorithms is the possibility of their incorporation into already available supercomputer simulation models of gas discs without need for alterations in their core methods of gas dynamics simulations. For example, this requirement is violated, if the transition to the conservative form of equations is required. A scheme for equations in a two-phase medium in a conservative form was developed by \citet{Miniati2010}. 

\section{Continuity and motion equations for the gas-dust medium of the circumstellar disc} 
\label{sec:model} 

In the present paper we consider a polytropic model for gaseous component of the disc. We assume that gas flow is described by two-velocity model (e.g. \citet{Marble1970,Nigmatullin}), where gas and dust exchange their momentum. Thus the continuity and motion equations for the dust and gaseous components have the following form: 

\begin{equation} 
\label{eq:gas} 
\displaystyle\frac{\partial \rho_{\rm g}}{\partial t}+\nabla (\rho_{\rm g} v)=0,\ \ \  
\displaystyle\frac{\partial v}{\partial t}+(v \cdot \nabla) v =-\frac{\nabla P}{ \rho_{\rm g}}+ g - \frac{K(v-u)}{\rho_{\rm g}}, 
\end{equation} 
\begin{equation} 
\label{eq:dust} 
\displaystyle\frac{\partial \rho_{\rm d}}{\partial t}+\nabla (\rho_{\rm d} u)=0,\ \ \  
\displaystyle\frac{\partial u}{\partial t}+(u \cdot \nabla) u = g + \frac{K(v-u)}{\rho_{\rm d}}, 
\end{equation} 
where $\rho_{\rm g}$ and $\rho_{\rm d}$ are volume densities of gas and dust, $v$ and $u$ are velocities of gas and dust, $P$ is the gas pressure, $g$ is the gravitational acceleration, $K(v-u)$ is the drag force between gas and dust. For the conditions of an circumstellar disc according to \citet{LaibePriceAstroDrag,Weidenschilling1977} the drag coefficient can assume various forms depending on the ratio between the dust grain size $s$ and mean free path of a molecule in the gas $\lambda$. For most parts of the disc (for detail estimations see, e.g. \citet{StoyanovskayaDust}) the Epstein drag mode of the interaction between gas and dust takes place, where 
\begin{equation} 
K \equiv \displaystyle\frac{\rho_{\rm d}}{t_{\rm stop}}, 
\end{equation} 
and $t_{\rm stop}$ is determined by (\ref{eq:t_stop_rho}). In this mode the drag force is a linear function of the relative velocity of gas and a solid body. Within the scope of the paper we consider only this drag mode, leaving non-linear Stokes regimes \citep{Weidenschilling1977} for further studies. 

The equation (\ref{eq:dust}) means, that the dynamics of the dust component of the disc is described in terms of a continuous medium with zero pressure and viscosity. The gas component of the disc occupies the entire available space, and the dust component has zero volume. This form of equation is a particular case of the dynamical equations for a gas-dust medium. In the other applications of the dynamics of gas and dust medium, for example, in the reactor simulations with pseudofluidized catalyst layer, the dust phase occupies a finite volume \citep{MonaghanKocharyan1995}, and the dust velocity is affected by the stress tensor in the dust component \citep{TFSPH}. 

\section{Computing drag force within TFSPH} 
\label{sec:DragINTFSPH} 

In the system (\ref{eq:gas})-(\ref{eq:dust}) the dust and gas are coupled via the drag force, depending on the relative velocity between gas and dust phases. If the velocities of dust and gas are known in the same points of space (grid methods, one-fluid smoothed particle hydrodynamics \citep{LaibePrice2014OneFluidDust}), then the relative velocity in these points can be determined uniquely. If gas and dust are modeled by different sets of particles, several methods are available for the computing of the drag force between the phases. 

\begin{figure*} 
\center
\includegraphics[width=0.9 \textwidth]{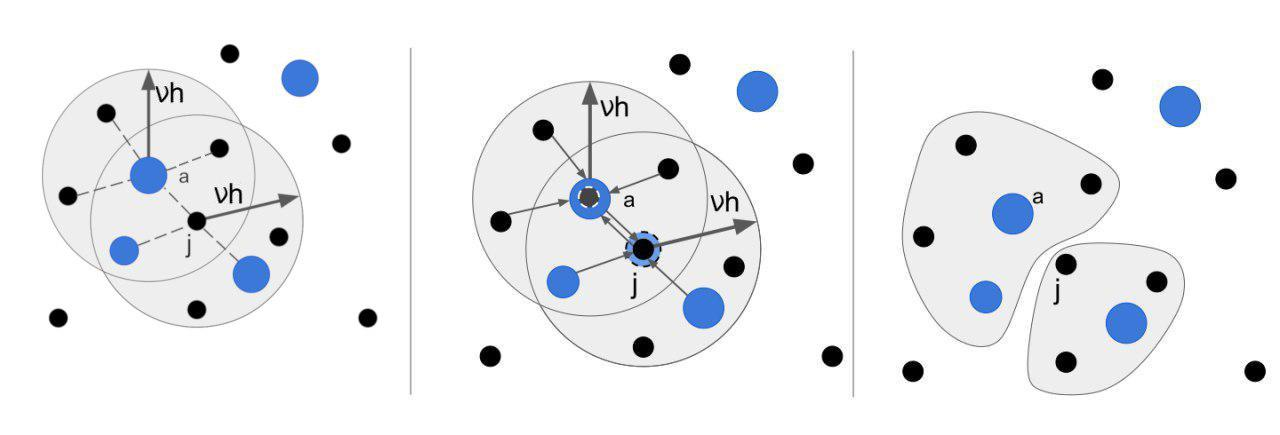} 
\caption{Principle of computing the drag force in two-fluid SPH: pair interaction of Monaghan-Kocharyan drag (MK, left panel), interpolation  (IPSPH, central panel) and averaging in Drag-in-Cell (DIC, right panel), $\nu h$ is a radius of kernel support domain, e.g. for cubic spline kernel (\ref{eq:kernel}) $\nu=2$. Dust particles are shown with small black circles, gas particles - with big blue circles. Dash lines indicate pairs in which drag force term is computed, using values of gas and dust particle, and than averaged with SPH interpolation technique. Arrows from particle to particle indicates interpolation of dust values into gaseous particle (and vice versa) before computing the local drag term.} 
\label{fig:SchemeScratch} 
\end{figure*}

A classical method for the calculation of the drag force within the smoothed particle hydrodynamics was introduced by \citet{MonaghanKocharyan1995}, further referred to as MKD (Monaghan-Kocharyan Drag). This method is based on the computing of the relative velocity between each pair of particles of gas and dust (Fig.\ref{fig:SchemeScratch}, left panel) and is used in the fundamental and engineering application of the two-fluid mechanics \citep{Maddison2004TFSPH,FranceDustCode,Gonzalez2017,TFSPH} and others. 

The other method is based on the calculation of gas density and velocity in the points where dust particles are located (and vice versa) with the use of SPH interpolation formulas. As a result, for every model particles all parameters of the gas and dust medium are known (Fig.\ref{fig:SchemeScratch}, central panel). Let us call this method IPSPH (InterPolation SPH). It was used in \citet{BateDust2014,ClarkeDust2015,RiceEtAl2004}. 

We propose to use the concept of the Particle-in-Cell method of simulation of gas-dust flows \citep{Andrews1996} for the computing of drag force. This concept is based on the idea of splitting the whole computational domain into nonoverlapping cells (a regular Cartesian or cylindrical coordinate grid can be used, but this is not a requirement) and calculating the averaged gas and dust velocities and densities within each cell. Next, the drag force affecting each particle can be calculated from the velocity of this particle and volume averaged parameters of the gas and dust medium (Fig.\ref{fig:SchemeScratch}, right panel). We will call this method DIC (Drag-In-Cell). 

In this article we compare MK and DIC methods from the point of view of their feasibility for the simulation of the coupled dynamics of gas and dust, interacting in the Epstein drag mode. Since the circumstellar may contains grains of varying sizes from 1 $\mu$m till few cm (upper boundary is debatable), we will focus on the relation between the method of drag force computing and the level of dissipation of the numerical schemes for media with arbitrary (high or low) drag coefficients. The comparison of MKD and IPSPH methods is given in \citet{BateDust2014}. 

\section{Compared schemes} 
\label{sec:methods} 

In this Section we specify the compared schemes for the solution of the equation of motion for gas and dust (\ref{eq:gas})-(\ref{eq:dust}) in the SPH standard notation. Let $n$ be the number of the time step. Following the convention, given in \citet{MonaghanKocharyan1995}, we shall use $a,b$ as indices for gas particles, and $j,k$ as indices for dust particles. For example, $\rho_{a, \rm g}, v_a, r_a, P_a$ denote density, velocity, coordinate and pressure of gas in the point with number $a$; $\rho_{j, \rm d}, u_{j}, r_j$ denote density, velocity and coordinates of the dust in the point with number $j$. 

\subsection{Explicit Monaghan-Kocharyan (MK) scheme} 

The classical SPH approximation for the equations of motion (\ref{eq:gas})-(\ref{eq:dust}) for the gas-dust medium \citep{MonaghanKocharyan1995} is based on the MKD method. We have implemented it in such a way, that the term, describing drag, uses the values of velocities from the previous time step:

\begin{align}
\displaystyle\frac{\mathrm{d}v^n_a}{\mathrm{d}t} &= - \sum_b m_{\rm g} \left(\frac{P_b}{(\rho^n_{b,\rm{g}})^2} + \frac{P_a}{(\rho^n_{a,\rm g})^2} + \Pi_{ab} \right) \bigtriangledown_a  W^{n}_{ab} - \nonumber 
\\& - \sigma m_{\rm d} \sum_j \frac{K_{aj}}{\rho^n_{a, \rm g} \rho^n_{j,\rm d}} \frac{(v_a^{n} - u_j^{n},r_{ja})}{r_{ja}^2+\eta^2}r_{ja}W^{n}_{ja} +g_a,
\label{eq:MonKoch1995v}
\end{align} 
\begin{equation} \displaystyle\frac{\mathrm{d}u^n_j}{\mathrm{d}t}= \sigma m_{\rm g} \sum_a \frac{K_{aj}}{\rho^n_{a, \rm g} \rho^n_{j,\rm d}} \frac{(v_a^{n} - u_j^{n},r_{ja})}{r_{ja}^2+\eta^2}r_{ja}W^{n}_{ja}+g_j, 
\label{eq:MonKoch1995u} 
\end{equation} 

\begin{equation} 
K_{aj}=\displaystyle\frac{\rho^n_{j,\rm d} \rho^n_{a, \rm g} c^n_{a,\rm s}}{s_j^n \rho^n_{j,\rm s}}, 
\end{equation} 
where $r_{ja} = r_j - r_a$, $\eta$ is a clipping constant, $\eta^2 = 0.001 h^2$ and $\sigma$ is a factor, determined by the dimensionality of the problem; for 1D problems $\sigma=1$. Here $\Pi_{ab}$ is the tensor of artificial viscosity, $W^{n}_{ab}=W(h,r_{ab})$ is the smoothing kernel.

\subsection{Novel SPH-IDIC scheme: implicit Drag-in-Cell} 

The second scheme is different from (\ref{eq:MonKoch1995v})-(\ref{eq:MonKoch1995u}) only by terms describing drag. 

For each time moment let us split the whole computational domain into nonoverlapping cells, so that the union of these sets coincides with the whole domain. Let a certain cell contains $N$ gas particles of the same mass $m_{\rm g}$ and $L$ dust particles with the same mass $m_{\rm d}$, with $N>0$, $L>0$. Let us introduce volume-averaged values of $t^*_{\rm stop}$ and $\rho^*_{\rm d}$ (averaging performed by any arbitrary method), and set 
\begin{equation} 
\label{eq:aveEpsilon} 
\varepsilon^*=\displaystyle\frac{m_{\rm d} L}{m_{\rm g} N}, 
\end{equation} 
thus defining 
\begin{equation} 
K^*=\frac{\rho^*_{\rm d}}{t^*_{\rm stop}}, \ \ \rho^*_{\rm g}=\frac{\rho^*_{\rm d}}{\varepsilon^*}. 
\end{equation}

We assume, that in the computing of the drag force, affecting the dust from the gas, the gas velocity is the same within the whole cell and equal to $v_*$, but the dust particles have different velocities (and vice versa). Besides that, we shall calculate the drag coefficient and densities, using calculated values from the previous time step, while the relative velocity will be used from the next step. The scheme thus has the following form: 

\begin{align} 
\displaystyle\frac{\mathrm{d}v^n_a}{\mathrm{d}t}= - \sum_b m_b & \left(\frac{P_b}{(\rho^n_{b,\rm{g}})^2} + \frac{P_a}{(\rho^n_{a,\rm g})^2} + \Pi_{ab} \right) \bigtriangledown_a W^{n}_{ab} - \nonumber \\ & -  \frac{K^*}{\rho^*_{\rm g}} (v_a^{n+1} - u_{*}^{n+1})+g_a, 
\label{eq:DragInCellv} 
\end{align}
\begin{equation} 
\displaystyle\frac{\mathrm{d}u^n_j}{\mathrm{d}t}= \frac{K^*}{\rho^*_{\rm d}} (v^{n+1}_* - u^{n+1}_j)+g_j, 
\label{eq:DragInCellu} 
\end{equation} 
\begin{equation} 
\label{eq:VUvolaverage} 
v_*=\displaystyle\frac{\sum_{i=1}^K v_i}{N}, \quad u_*=\displaystyle\frac{\sum_{j=1}^L u_j}{L}. 
\end{equation}

If the time derivative in (\ref{eq:DragInCellv})-(\ref{eq:DragInCellu}) is approximated by the first order finite difference, then a fast way for computing of $u^{n+1}$, $v^{n+1}$ can be proposed (see \ref{sec:implicitdrag}). 

In the next Section we describe the test problems to be used for comparison of the approaches (\ref{eq:MonKoch1995v})-(\ref{eq:MonKoch1995u}) and (\ref{eq:DragInCellv})-(\ref{eq:DragInCellu}).

\section{Test 1. Dusty Wave - sound waves in a two-phase isothermal medium} 
\label{sec:DustyWave} 

\subsection{Problem statement. Reference solution} 
For an isothermal medium the system of equation  (\ref{eq:gas})-(\ref{eq:dust}) takes the following form: 

\begin{equation} 
\label{eq:DustyWaveCont} 
\frac{\partial \rho_{\rm g}}{\partial t}+\frac{\partial{(\rho_{\rm g} v)}}{\partial x} = 0, \ \  \frac{\partial \rho_{\rm d}}{\partial t}+\frac{\partial{(\rho_{\rm d} u)}}{\partial x} = 0,\ \  
\end{equation}

\begin{equation} 
\label{eq:DustyWaveMotionGas} 
\rho_{\rm g} (\frac{\partial v}{\partial t}+v \frac{\partial v}{\partial x}) = - c_s^2 \frac{\partial \rho_{\rm g}}{\partial x} - K(v-u), 
\end{equation} 

\begin{equation} 
\label{eq:DustyWaveMotionDust} 
\rho_{\rm d} (\frac{\partial u}{\partial t}+u \frac{\partial u}{\partial x}) = K(v-u). 
\end{equation}

The steady solution of the system (\ref{eq:DustyWaveCont})-(\ref{eq:DustyWaveMotionDust}) is given by the functions 
\begin{align} 
\label{eq:SteadySolution} 
\rho_{\rm g}(x)=\tilde{\rho_{\rm g}}=const, \ \ \rho_{\rm d}(x)=\tilde{\rho_{\rm g}}=const&, \ \  v(x)=0, \nonumber \\ & u(x)=0. 
\end{align} 

Consider the solution to the system (\ref{eq:DustyWaveCont})-(\ref{eq:DustyWaveMotionDust}) on the interval $x \in [0,1]$ with positive sound speed, setting periodic conditions by $x$ for the solutions on the left border: 
\begin{align} 
\label{eq:periodic} 
\rho_{\rm g}|_{x=0}=\rho_{\rm g}|_{x=1}, \ \ \rho_{\rm d}|_{x=0}=\rho_{\rm d}|_{x=1}, \ \ &v|_{x=0}=v|_{x=1}, \nonumber \\ &u|_{x=0}=u|_{x=1}, 
\end{align}
and initial data as small perturbations of stationary density and velocity (\ref{eq:SteadySolution}): 
\begin{equation} 
\label{eq:DustyWave_init1} 
\rho_{\rm g}|_{t=0}=\tilde{\rho_{\rm g}}+A \sin(kx), \ \rho_{\rm d}|_{t=0}=\tilde{\rho_{\rm d}}+A \sin(kx), 
\end{equation} 
\begin{equation} 
\label{eq:DustyWave_init2} 
v|_{t=0}=A \sin(kx), \ u|_{t=0}=A \sin(kx). 
\end{equation} 
Here $k$ is wave number, defining the integer number of sine waves of density and velocity on the interval $[0,1]$, $A$ is the perturbation amplitude. In the vicinity of (\ref{eq:SteadySolution}) the linearized system (\ref{eq:DustyWaveCont})-(\ref{eq:DustyWaveMotionDust}) for perturbations has the form:

\begin{equation} 
\label{eq:LinDustyWaveCont} 
\frac{\partial (\delta \rho_{\rm g})}{\partial t}+\tilde{\rho_{\rm g}} \frac{\partial{v}}{\partial x} = 0, \ \  \frac{\partial (\delta \rho_{\rm d})}{\partial t}+\tilde{\rho_{\rm d}} \frac{\partial{ u}}{\partial x} = 0,\ \  
\end{equation}

\begin{equation} 
\label{eq:LinDustyWaveMotionGas} 
\tilde {\rho_{\rm g}} \frac{\partial v}{\partial t} = - c_s^2 \frac{\partial (\delta \rho_{\rm g})}{\partial x} - K(v-u), 
\end{equation} 

\begin{equation} 
\label{eq:LinDustyWaveMotionDust} 
\tilde {\rho_{\rm d}} \frac{\partial u}{\partial t} = K(v-u). 
\end{equation}

The reference solution of the linearized system (\ref{eq:LinDustyWaveCont})-(\ref{eq:LinDustyWaveMotionDust}) is given in \citet{LaibePrice2011}, who also shared the code for generation of this solution, which we used here. 

For simplicity, in what follows we shall refer to the reference solution of the linearized system (\ref{eq:LinDustyWaveCont})-(\ref{eq:LinDustyWaveMotionDust}) as the analytical solution of DustyWave problem.

The linearized system (\ref{eq:LinDustyWaveCont})-(\ref{eq:LinDustyWaveMotionDust}) has an analytical solution for dusty gas mixtures with both small dust grains (strong drag) and large bodies (weak drag). This key feature of the problem allows it to be used for assessment of the applicability of the method for simulation of a solid phase with arbitrary size of particles. 

\subsection{Numerical algorithm, boundary and initial conditions}

The continuity equations for dust and gas are approximated in a standard way within SPH: 
\begin{equation} 
\label{eq:contin_gas} 
\rho^{n}_{a, \rm g} = m_{\rm g} \sum_b W^{n}_{ab}, 
\end{equation} 
\begin{equation} 
\label{eq:contin_dust} 
\rho^{n}_{i, \rm d} = m_{\rm j} \sum_i W^{n}_{ij}. 
\end{equation}

Most of test runs are done with cubic spline kernel, that is classical for SPH and used in astrophysical (e.g.\cite{FranceDustCode,Gonzalez2017}) and industrial (e.g.\cite{TFSPH}) simulations:

\begin{align} 
\label{eq:kernel} 
W^{n}_{ab} =& W(|r^{n}_a - r^{n}_b|, h) =  \nonumber \\  = & W^{n}(q) = \frac{2}{3 h} 
\begin{cases} 1 - \frac{3}{2} q^2 + \frac{3}{4} q^3, &\text{if $ \displaystyle 0 \le q \le 1$,}\\ 
\displaystyle \frac{1}{4}(2 - q)^3, &\text{if $\displaystyle 1 \le q \le 2$,}\\ 0, &\text{otherwise;} 
\end{cases} 
\end{align} 
where $q = \displaystyle \frac{|r^{n}_a - r^{n}_b|}{h}$.

However, recent works by \cite{LaibePrice2011Test,BateDust2014} underlined that quintic kernel increases the accuracy of drag terms computing. For this reason in Section \ref{sec:kernels} we compared cubic spline and quintic kernel results. We used quintic kernel
\begin{align} 
\label{eq:quintic_kernel} 
& W^{n}(q) = \nonumber \\ & = \frac{\sigma}{h} 
\begin{cases} \displaystyle (3\xi-q)^5 - 6(2\xi-q)^5 + 15(1\xi-q)^5, &\text{if $ \displaystyle 0 \le q < \xi$,}\\ 
\displaystyle (3\xi-q)^5 - 6(2\xi-q)^5, &\text{if $\displaystyle \xi \le q < 2\xi$,}\\
\displaystyle (3\xi-q)^5, &\text{if $\displaystyle 2\xi \le q < 3\xi$,}\\ 0, &\text{otherwise;} 
\end{cases} 
\end{align} 
with radius of support domain $h$ ($\xi = \displaystyle \frac{1}{3}, \sigma = \frac{3^5}{40}$) \citep{Wendland} and $3h$ ($\xi = 1, \displaystyle \sigma = \frac{1}{120}$)\citep{Price}.

In the spatial points, occupied by gas particles, the pressure is calculated as $P_a = c^2_s \rho_{a, \rm g}$. The artificial viscosity is not introduced in this test problem, that is $\Pi_{ab} = 0$. 

The time integration of the equations of motion (\ref{eq:MonKoch1995v})-(\ref{eq:MonKoch1995u}), (\ref{eq:DragInCellv})-(\ref{eq:DragInCellu}) is performed with an explicit first order approximation scheme. The time step is determined from the Courant condition: 
\begin{equation} 
\label{eq:courant} 
\tau < \displaystyle \frac{h \cdot CFL}{\textrm{max}(c_s, u, v)}, 
\end{equation} 
where $CFL$ is the Courant parameter.

The initial distributions of gas and dust velocities are set as $u_0 = v_0 = \displaystyle A \sin(2\pi x)$, gas density $\rho_{\rm g,0} = \displaystyle A \sin(2\pi x) + 1$, dust density $\rho_{\rm d,0} = \displaystyle A \sin(2\pi x) + \varepsilon$. Following parameters were used in the calculations: $A = 10^{-4}$, $c_s = 1$, $\varepsilon = 1$. 

To obtain the density on the interval $[0,l]$, which is perturbed around the constant value $\tilde{\rho}$ by the value $A \sin(2\pi x)$, that is $\rho_0(x)=\tilde{\rho}+A \sin(2\pi x)$, we used the recurrent procedure of the model particles placement. The first particle is placed into the origin of the coordinates $x_1=0$, and the coordinate of the next particle is calculated from the relation 

$$\displaystyle \int\limits_{x_i}^{x_i + \Delta x_i} \rho_0 dx = \frac{\tilde{\rho} l}{N_{\rm ph}},$$ where $N_{\rm ph}$ is the number of model gas or dust particles. After all $x_i$ are determined, each particle is shifted to the right by a value $\Delta x_i$.

In the initial time moment the particles are located within the interval [0,1], and the coordinates of gas and dust particles coincide. To ensure the periodic boundary conditions (\ref{eq:periodic}) for each time step, the particles from the interval were cloned with placement them at a distance, equal to the length of the interval, to the left and to the right. 

If not otherwise specified, the calculations use an equal number of gas and dust particles $N_{\rm total} = 2 \times 600$, a constant smoothing length $h=0.01$ and the time step $\tau = 0.001$, ensuring $CFL=0.1$. 

\section{Test 2. Dustyshock - shock tube problem for gas and dust medium} 
\label{sec:DustyShock}

\subsection{Problem statement. Reference solution} 

In this Section we consider the shock tube problem as a classical test for gas dynamics solvers, which is often referenced as Sod test \citep{Sod1978}. The modification of this problem is also often used for testing computation schemes for two-phase media, for example, \citet{LaibePrice2011Test,Saito2003}. The one-dimensional equations for conservation of mass, momentum and energy in the gas-dust medium have the following form in the notations of Section \ref{sec:model}:

\begin{equation} 
\label{eq:ShockWaveCont} 
\frac{\partial \rho_{\rm g}}{\partial t}+\frac{\partial{(\rho_{\rm g} v)}}{\partial x} = 0, \ \  
\frac{\partial \rho_{\rm d}}{\partial t}+\frac{\partial{(\rho_{\rm d} u)}}{\partial x} = 0,\ \  
\end{equation}

\begin{equation} 
\label{eq:ShockWaveMotionGas} 
\rho_{\rm g} (\frac{\partial v}{\partial t}+v \frac{\partial v}{\partial x}) = - \frac{\partial P}{\partial x} - K(v-u), 
\end{equation} 

\begin{equation} 
\label{eq:ShockWaveMotionDust} 
\rho_{\rm d} (\frac{\partial u}{\partial t}+u \frac{\partial u}{\partial x}) = K(v-u), 
\end{equation} 

\begin{equation} 
\label{eq:ShockWaveEnergyGas} 
\rho_{\rm g} (\frac{\partial e}{\partial t} +v\frac{\partial e}{\partial x})=-P\frac{\partial v}{\partial x}, 
\end{equation} 
where $e$ is the internal energy (temperature) of the gas, related in the following way with its pressure: 
\begin{equation} 
\label{eq:EOS} P=\rho_{\rm g} e (\gamma-1). 
\end{equation}

For the system (\ref{eq:ShockWaveCont})-(\ref{eq:ShockWaveEnergyGas}) the flow conditions are set on the boundaries of the interval. Zero initial velocities, gas pressure jump, and density jumps for gas and dust are imposed at the initial time moment. If a solid phase is absent in the gas, then the analytical solution of the problem is known for the whole range of parameters. The reference solution for the gas-dust medium is known for the steady case, that is for time moments $t>t_{\rm stop}$ under the condition, that $t_{\rm stop} \max(c_s,u,v) \ll l$, where $l$ is the linear size of the computational domain. This solution is obtained from the solution for gas dynamics by replacement of the sound speed in the gas with the sound speed in the gas-dust medium (e.g., \citet{LaibePrice2011Test}): 
\begin{equation} 
\label{eq:dustysound} 
c^*_s=c_s(1+\displaystyle\frac{\rho_{\rm d}}{\rho_{\rm g}})^{-1/2}. 
\end{equation}

We shall consider the case, when the drag term in the r.h.s. of (\ref{eq:ShockWaveMotionGas}) essentially exceeds the term with the pressure gradient. This allows to exclude the quadratic term $(u-v)^2$ from consideration in the equation (\ref{eq:ShockWaveEnergyGas}), since it is of the second order of smallness with respect to $(u-v)$. 

\subsection{Numerical algorithm, boundary and initial conditions} 

The continuity conditions are approximated with equations (\ref{eq:contin_gas})-(\ref{eq:contin_dust}), the kernel is chosen in the form (\ref{eq:kernel}), as in the case of the DustyWave problem. 

The SPH approximation of the energy equation with artificial viscosity are taken in the form: 
\begin{equation} 
\label{eq:energy} 
\displaystyle \frac{\mathrm{d}e^n_a}{\mathrm{d}t} = \frac{m_{\rm g} P_a}{(\rho^n_{\rm g, a})^2} \sum_b (v^n_a - v^n_b) \nabla_a W^n_{ab} + \frac{m_{\rm g}}{2} \sum_b \Pi_{ab} (v^n_a - v^n_b) \nabla_a W^n_{ab}, 
\end{equation}

where 
\begin{equation} 
\Pi_{ab} = 
\begin{cases} \displaystyle \frac{-\alpha c_{ab} \mu_{ab} + \beta \mu^2_{ab}}{\rho_{ab}}, &\text{if $ v_{ab} r_{ab} < 0$,}\\ 0, &\text{if $ v_{ab} r_{ab} > 0$,} \end{cases} 
\end{equation} 
$\mu_{ab} = \displaystyle \frac{h v_{ab} r_{ab}}{r^2_{ab} + \nu^2},$ \ \ $v_{ab} = v_a - v_b$, \ \ $r_{ab} = r_a - r_b$, \ \ \\ $\rho_{ab} = \displaystyle \frac{1}{2} (\rho_a + \rho_b)$, \ \ $c_{ab} = \displaystyle \frac{1}{2} (c_a + c_b)$ \ \ and \ \ $c = \sqrt[]{\displaystyle \frac{\gamma P}{\rho}}$.

We have used standard parameters of the artificial viscosity $\alpha = 1, \beta = 2, \nu = 0.1 h$ \citep{SPH}. 

The time integration of the equations of motion (\ref{eq:MonKoch1995v})-(\ref{eq:MonKoch1995u}), (\ref{eq:DragInCellv})-(\ref{eq:DragInCellu}) and energy (\ref{eq:energy}) is performed with an explicit first order approximation scheme. 

As initial data we set the values for the left and right domains at zero time: $$[\rho_{l}, P_l, v_l, e_l] = [1, 1, 0, 2.5],$$ $$[\rho_r, P_r, v_r, e_r] = [0.125, 0.1, 0, 2],$$ $$\gamma_l = \gamma_r = 7/5.$$ The initial dust to gas mass ratio is set $\varepsilon=1$. If not otherwise specified, 990 gas and 990 dust particles is used: $N_{\rm total} = 2 \times 990$. In the initial time moment 880 particles are homogeneously distributed on the interval $[-0.5, 0]$, and 110 particles - on the interval $[0, 0.5]$. Dust particles are placed in the same points as gas particles. 

To mimic the boundary conditions, $450$ immobile particles are placed outside the interval [-0.5,0.5] \citep{Morris1997}, which ensure a close to zero pressure gradient and zero density drops at the ends of the interval.

If not otherwise specified, the calculations use a constant smoothing length $h=0.01$ and time step $\tau = 0.001$, satisfying the Courant condition with $CFL=0.1$. 

\section{Results and discussion} 
\label{sec:resultsDiscussion}

In this Section we compare the numerical solutions, obtained from MK and SPH-IDIC methods for different values of drag coefficient (from weak to stiff drag). 

For the cases when reference solution of the problem is available, we quantify the scheme error using $L_2$ norm:

\begin{equation}
\label{eq:L2norm}
L_2 = \left[ \frac{1}{N_{\rm ref}} \left(  \sum^{N_{\rm ref}}_{i=1} \frac {(f_i - f_{\rm exact})^2}{f^2_{\rm max}} \right) \right]^{1/2},
\end{equation}

\noindent where $f_{\rm max}$ - maximum of reference solution on computational domain, $f_{\rm exact}$ is the reference solution for the $i$-th point, $N_{\rm ref}$ - number of reference solution points. We compute $f_i$ in reference solution points, using standard SPH interpolation $\displaystyle f_i = m \sum_j \frac{f_j}{\rho_j} W_{ij}$ (where $j$ - number of SPH particle). We note however, that reference solution is not exact analytical solution, but its approximation with unknown error, so the obtained from (\ref{eq:L2norm}) value is useful approximation of scheme error.

For the implementation of the SPH-IDIC scheme the expressions from \ref{sec:implicitdrag} were used. To calculate the drag force, the whole computational domain is split into equal sections with length $h_{\rm cell}$. If not otherwise specified, $h_{\rm cell}=h$. For each cell, $\rho^*_{\rm d}$ was calculated as $\rho^*_{\rm d} = \displaystyle \frac{\sum_{j=1}^L \rho_{\rm d}}{L}$, $t^*_{\rm stop} = \displaystyle \frac{\rho^*_{\rm d}}{K}$. The MK scheme was implemented with a constant drag coefficient $K$. 

The results presented in this Section were obtained from running our developed codes, written is C language. To download the codes please use the links

https://bitbucket.org/astrosolvers/dustywave/downloads/, 

https://bitbucket.org/astrosolvers/dustyshock/downloads/

(available for public access) or clone the corresponding git repositories. The codes were developed to test the features of different schemes on one-dimensional problems, however, it were not specially optimized by performance.

\subsection{Comparison of schemes on problems with weak and moderate drag} 
\label{sec:mild} 

\begin{figure*}
\center
\includegraphics[width=0.8 \textwidth]{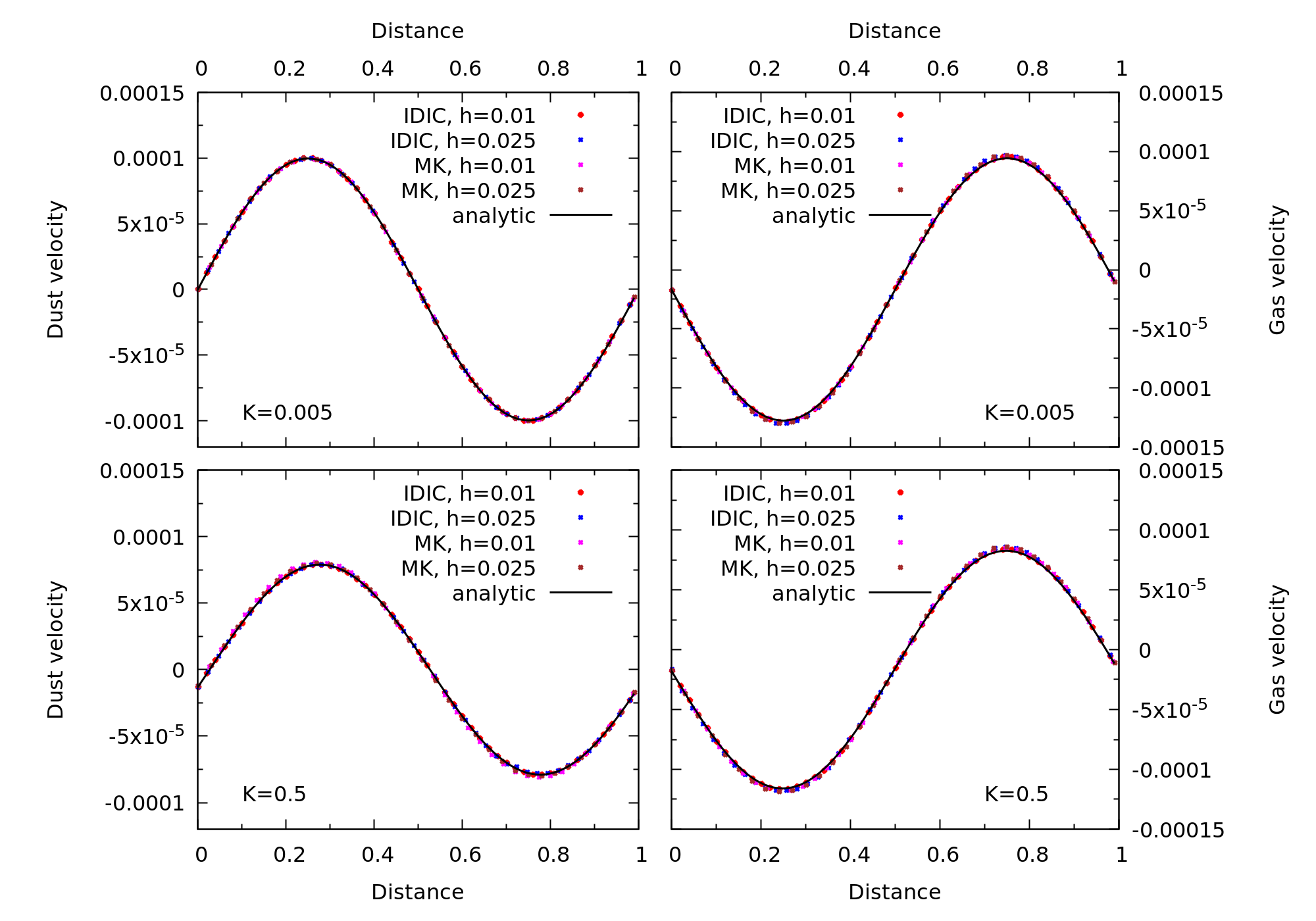}
\caption{Solution of the DustyWave problem for the time moment $t=0.5$. Comparison of MK and SPH-IDIC scheme results for weak and moderate drag. Top panels: $K=0.005$, bottom panels: $K=0.5$. Left panels present the dust velocity, right panels are for gas velocity. Total number of particles, involved in the computation: $N_{\rm total}=2 \times 600$, $CFL=0.1$, cubic kernel (\ref{eq:kernel}) $M_4$ is implemented.} 
\label{fig:Waved2g1_Ksmall_4frame} 
\end{figure*}

Primarily, we compared the results obtained with the MK and SPH-IDIC schemes under the same spatial resolution, when $t_{\rm stop}$ is comparable to or much greater than the dynamical time scale of the problem. In Fig.\ref{fig:Waved2g1_Ksmall_4frame} we present the analytical and numerical solutions to the DustyWave problem at the time moment $t=0.5$ for $K=0.005$ (top panels) and $K=0.5$ (bottom panels). Left panels present the dust velocity, right panels are for gas velocity. In both cases the numerical solution is close to the analytic one. In the mode with practically no coupling between the gas and dust ($K=0.005$), the dust velocity at $t=0.5$ is close to the initial value. Moreover the perturbation in the gas has shifted to the right with the sound speed $c_s$, (rather than $c^*_s$ from (\ref{eq:dustysound})), and its amplitude increased. For a moderate drag $K=0.5$ compared to the weak drag with $K=0.005$, wave damping occurred for dust velocities, which follows from the disperse relation from \citet{LaibePrice2011Test} and coincides with the results of \citet{LaibePrice2011}.

To quantify the accuracy of the DustyWave results in Tab.\ref{tab:DuSTYWAVEweakmod} we provide numerical solution error for dust velocity, that was found using $L_2$ norm (\ref{eq:L2norm}). Inspection of Tab.~\ref{tab:DuSTYWAVEweakmod} reveals that in the case of weak drag $K=0.005$ MK and IDIC schemes provides accurate results (better than 1~\%) with almost similar solution error. For this case the error demonstrates the second order dependence on $h$. However, for the case of moderate drag $K=0.5$ the solution error for IDIC scheme is much less than for MK. Moreover, error for IDIC scheme demonstrates second order dependence on $h$, while for MKD scheme it is only first order dependent. 

\begin{table}
\caption{$L_2$ error of dust velocity for DUSTYWAVE problem results from Fig.~\ref{fig:Waved2g1_Ksmall_4frame} with weak and moderate drag. As with data plotted on Fig.~\ref{fig:Waved2g1_Ksmall_4frame} cubic kernel is implemented, $N_{\rm total}=2 \times 600$ particles involved in the computation.}
\begin{center}
\begin{tabular}{|c|c|c|c|c|}
\hline & & & K=0.005 & K=0.5 \\
\hline \multirow{2}{*}{$h=0.01$} & $\tau=0.001$ &MK & 0.000636 & 0.012058 \\ & $\tau=0.001$ &IDIC & 0.000653 & 0.001237 \\ 
\hline \multirow{2}{*}{$h=0.025$} & $\tau=0.0025$ &MK & 0.002829 & 0.024018  \\ & $\tau=0.0025$ &IDIC & 0.002954 & 0.004631 \\ 
\hline
\end{tabular}
\label{tab:DuSTYWAVEweakmod}
\end{center}
\end{table}

\begin{figure*}
\includegraphics[width=0.8 \textwidth]{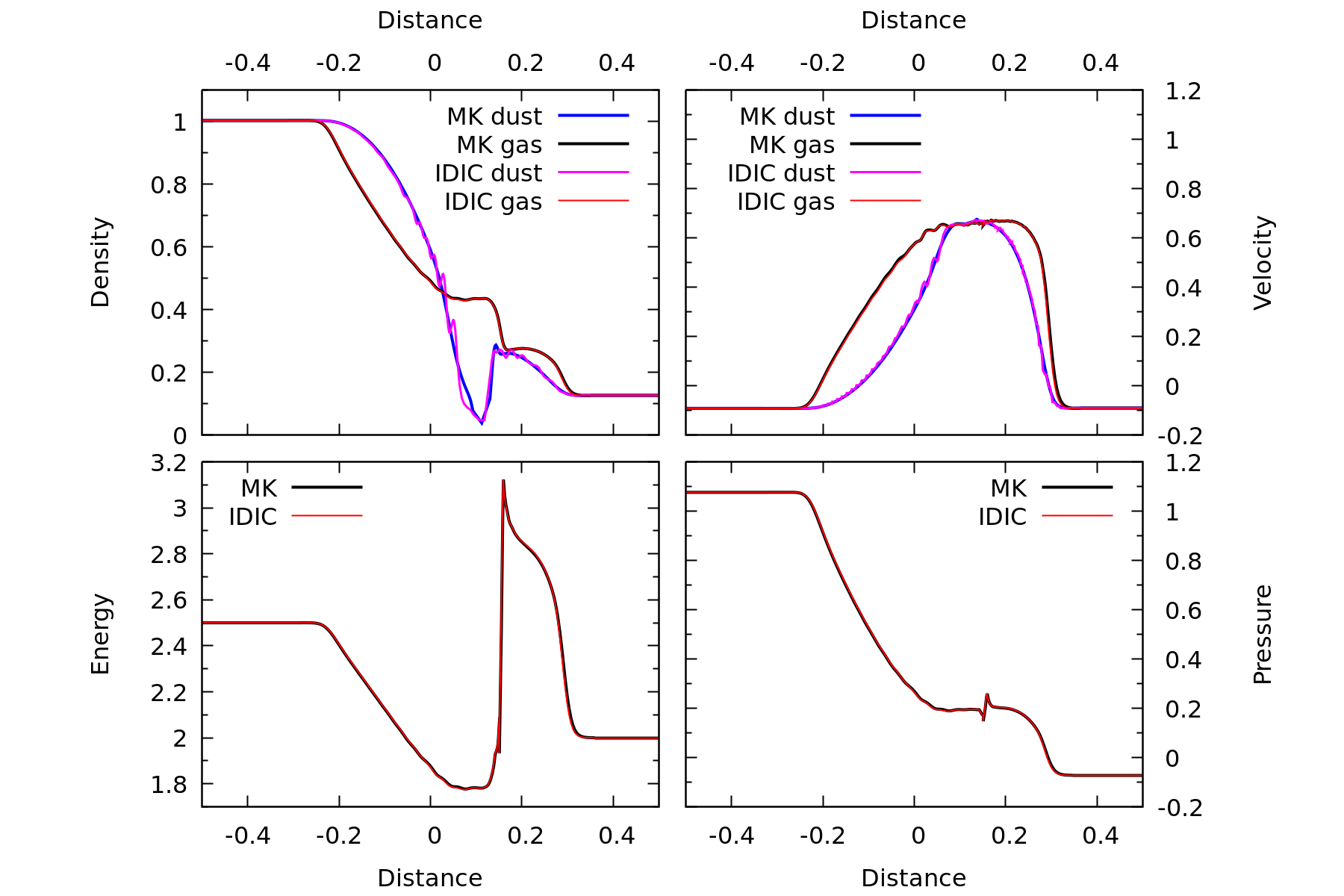} 
\caption{Solution of the DustyShock problem for the time $t=0.2$. Comparison of MK and SPH-IDIC scheme results at moderate drag $(K=5)$. Upper panels: densities and velocities for gas and dust, lower panels: gas energy and pressure. Total number of particles, involved in the computation $N_{\rm total}=2 \times 990$, $h=0.01$, $\tau=0.001$, cubic kernel (\ref{eq:kernel}) $M_4$ is implemented.} 
\label{fig:d2g1_K5_monaghan+cells} 
\end{figure*}

In Fig.\ref{fig:d2g1_K5_monaghan+cells} the numerical solution for the DustyShock problem with $K=5$ is given. The reference solution for this case of a moderate drag is not available. Both schemes yield similar results for gas and dust velocities, gas density, internal energy and pressure. Moreover, the obtained numerical solutions qualitatively coincide with the solution obtained by \citet{LaibePrice2011} (see Fig.11 in their paper). However, in the vicinity of the contact discontinuity, the MK scheme reproduces a more smoothed dust density compared to the SPH-IDIC scheme. 

Therefore, for the considered values of drag coefficient (weak and moderate drag) the smoothing length satisfies the condition (\ref{eq:sparesSPH}), and the time step, chosen from the Courant condition (\ref{eq:courant}) satisfies the condition  (\ref{eq:timeresSPH}). The Fig.\ref{fig:Waved2g1_Ksmall_4frame} and Fig.\ref{fig:d2g1_K5_monaghan+cells} show, that the results obtained with both schemes, with similar numerical resolution, are comparable in accuracy.

\subsection{Comparison of schemes on problems with stiff drag} 
\label{sec:stiff} 

\begin{figure*}[h]
\center
\includegraphics[width=0.9 \textwidth]{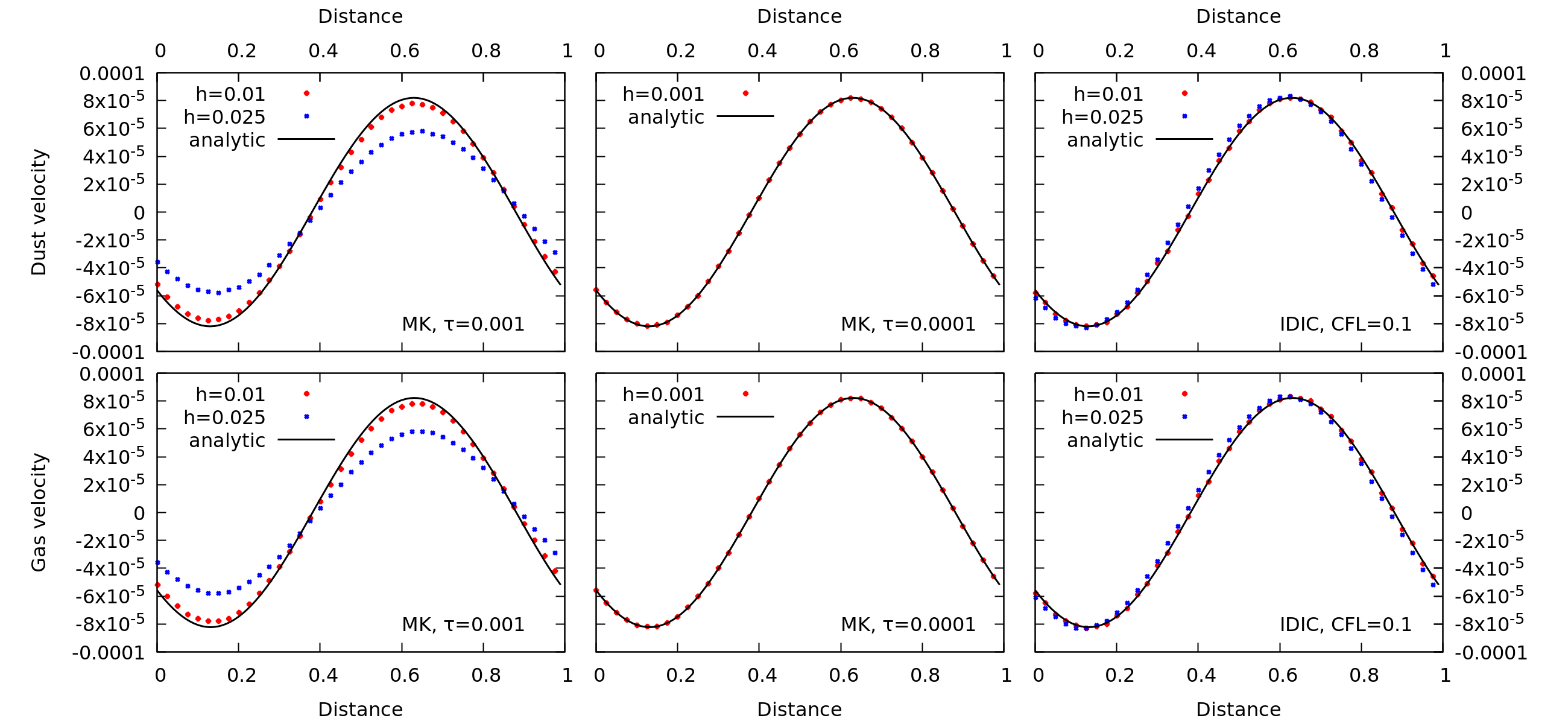} 
\caption{Solution of the DustyWave problem for the time moment $t=0.5$. Comparison of the MK and SPH-IDIC scheme results for stiff drag $K=500$, $\varepsilon=1$, cubic kernel is used. Top panels show dust velocity, bottom panels - gas velocity. Two left panels show the MK scheme with $h=0.01$, violating the condition (\ref{eq:sparesSPH}), central panels correspond to $h=0.001$, satisfying the condition (\ref{eq:sparesSPH}). Two right panels show the SPH-IDIC scheme with $h=0.01$, violating the condition (\ref{eq:sparesSPH}). Left and right panels -- $2\times 600$ SPH particles is used, center panels -- $2\times 6000$ particles is used.} 
\label{fig:T0.5_wave_d2g1_K500} 
\end{figure*}

At the second stage, we compare the numerical solutions obtained with the MK and SPH-IDIC schemes under the same spatial resolution, when $t_{\rm stop}$ is much less then dynamical time of the problem. Four left panels of the Fig.\ref{fig:T0.5_wave_d2g1_K500} show the analytical and numerical solution to the DustyWave problem at $K=500$ for the time moment $t=0.5$. Top panels show dust velocity, bottom panels - gas velocity. It can be seen, that in a mixture with stiff drag, velocities of gas and dust coincide by both phase and amplitude. Left panels show the calculation results with a fixed step $\tau=0.001$ at different smoothing lengths. The bigger is the smoothing length, the stronger is the dissipation of the velocities. The comparison of Fig.\ref{fig:Waved2g1_Ksmall_4frame} and Fig.\ref{fig:T0.5_wave_d2g1_K500} shows, that the level of dissipation depends on the drag parameters. Thus, at $h=0.025$ and $K \leq 0.5$ it is not noticeable, however, for the same $h$ and $K=500$ it has a pronounced manifestation. The central panels show the results for such calculation with increased number of particles $N_{\rm total}=6000$ and decreased smoothing length $h=0.001$. Now enhanced spatial resolution satisfies (\ref{eq:sparesSPH}). Obviously, the computation with such an enhanced resolution is free of the dissipation of gas and dust velocities.

\begin{figure*}[h]
\center
\includegraphics[width=0.9 \textwidth]{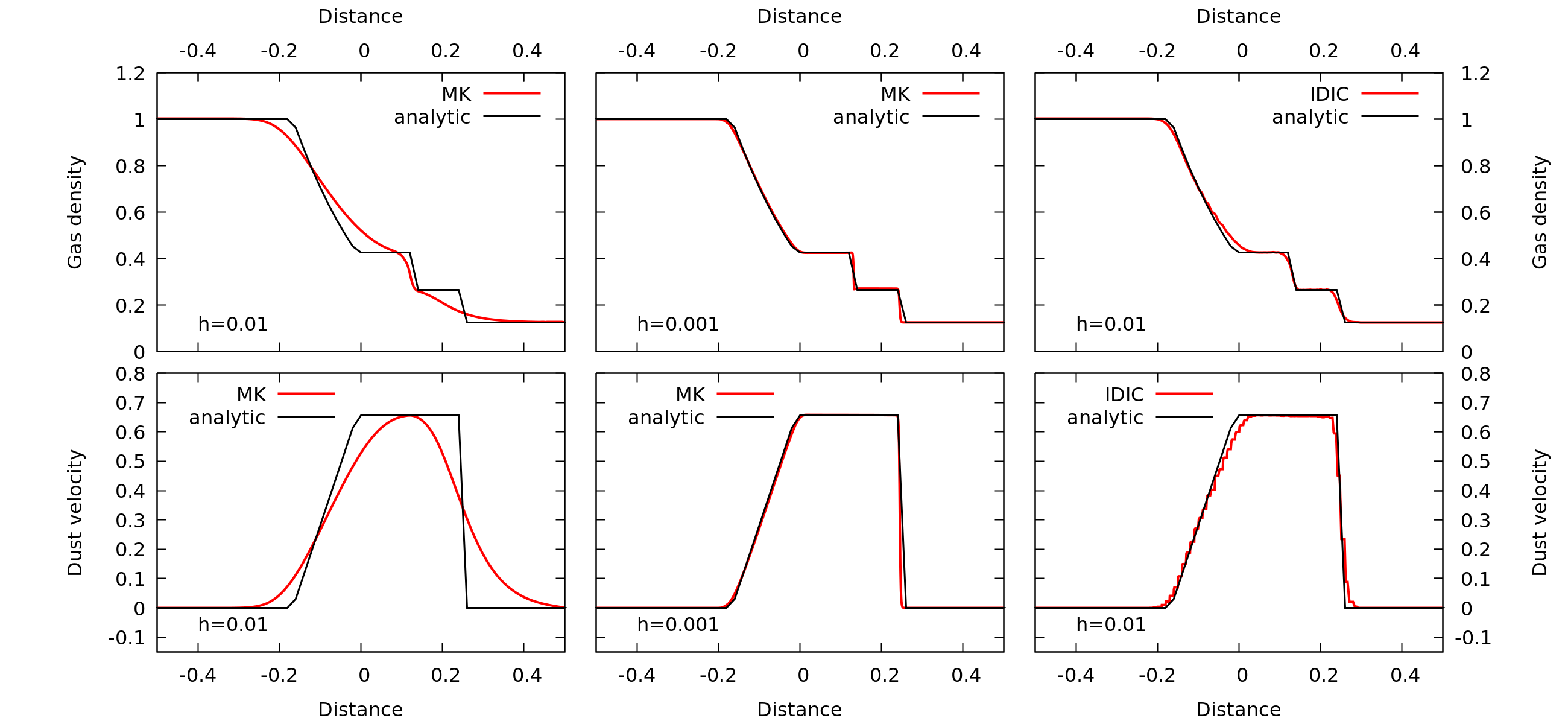} 
\caption{Solution of the DustyShock problem for the time $t=0.2$. Comparison of the MK and SPH-IDIC schemes at $t_{stop}=0.002$, $K=500$, $\varepsilon=1$. Top panels show gas density, bottom panels - dust velocity. Two left panels show the MK scheme with $h$, violating the condition (\ref{eq:sparesSPH}), central panels correspond to $h$, satisfying the condition (\ref{eq:sparesSPH}). Two right panels show the SPH-IDIC scheme with $h$, violating the condition (\ref{eq:sparesSPH}). On the left panels $h=0.01, \tau=10^{-3}, 2\times 990$ SPH particles, on the central panels $h=0.001, \tau = 10^{-4}, 2\times 9900$ SPH particles, on the right panels $h=0.001, \tau=10^{-3}, 2\times 990$ SPH particles.} 
\label{fig:d2g1_K500_shock} 
\end{figure*}

The similar effect occurs for the numerical solution of the DustyShock problem as well. Four left panels in Fig.\ref{fig:d2g1_K500_shock} present the numerical solution of the DustyShock problem for the time moment $t=0.2$ for stiff drag with $K=500$. In this case, $t_{\rm stop}=0.002 \ll t$, therefore to estimate the accuracy of the numerical solution, we can use a reference solution with $c^*_s$ for a gas-dust medium (\ref{eq:dustysound}). Two left panels show the numerical solution with $h=0.01$, violating the condition (\ref{eq:sparesSPH}). It is seen, that for both dust velocity and gas density an over-dissipation is observed. As can be noted from the bottom left panel, the scale of dissipation  for the solution exceeds 10 times the smoothing length $h$. We note that for the shock tube problem in gas with the same values $h$ and $\tau$ such a dissipation is not observed (see Fig.\ref{fig:d2g1_K5000_cells} and Section \ref{sec:Cells_DustyGasVSGas}). When the smoothing length was decreased to $h=0.001$ (the number of model particles was increased 10 times), the dissipation disappeared.

Right panels of Fig.\ref{fig:d2g1_K500_shock} show the results of the solution of the same problem with the SPH-IDIC scheme. The step $h$ is selected to be the same, as for the calculations with MK on the left panels of Fig.\ref{fig:d2g1_K500_shock}, it is seen, that for stiff drag, under similar spatial resolution, the results of the SPH-IDIC scheme essentially differ from those within the MK scheme. An over-dissipation, observed in the MK scheme, is absent in the calculations according to the SPH-IDIC scheme, however, a weak nonmonotonicity in the dust velocity appears. 

The absence of over-dissipation in the SPH-IDIC scheme also follows from the right panels of Fig.\ref{fig:T0.5_wave_d2g1_K500}, where the numerical solutions to the DustyWave problem with the smoothing length $h=0.01$ and $h=0.025$ are shown. Comparing left and right panels of Fig.\ref{fig:T0.5_wave_d2g1_K500} we see, that at $h=0.025$ the wave amplitude from the MK scheme is 30\% less, than the reference value. Meanwhile, the error within the SPH-IDIC scheme does not exceed several percents. We underline also, that the numerical solution at $h=0.025$ was obtained with $\tau>t_{\rm stop}$ due to the use of implicit drag approximation in SPH-IDIC.

Thus, we conclude, that for all tests the numerical solutions obtained with the MK scheme yield an over-dissipation for mixtures with high dust concentration and stiff coupling between the gas and the dust. This coincides with the results of \citet{LaibePrice2011} (see Fig.~8 and Fig.~12 of their article) and \citet{BateDust2014}. On the other hand, the SPH-IDIC scheme is free from that over-dissipation when similar spatial resolution is used.

\subsection{Dustyshock in gas and dust mixture with extremely stiff drag, notes on cell size and smoothing length relations} 
\label{sec:Cells_DustyGasVSGas} 

In Section \ref{sec:stiff} we have shown, that for mixtures with high dust concentration and stiff drag the accuracy of the solution by the MK scheme depends crucially on the smoothing length. Therein, the design of the method allows for use of different smoothing lengths for the approximation of the pressure gradient and drag force in (\ref{eq:MonKoch1995v})-(\ref{eq:MonKoch1995u}). Besides that, \citet{LaibePrice2011Test} proposed to use different kernels in the calculation of the pressure and drag forces. In particular, they achieved an increase in accuracy of the method, taking a quintic or double-humped kernel for the drag force \citep{FUlkQuinn1996}.

The design of the SPH-IDIC method also allows for different values of the smoothing length and typical cell size $h_{\rm cell}$, in which the drag is calculated. Varying $h$ and $h_{\rm cell}$ in the SPH-IDIC scheme we revealed, that for mixture with high dust concentration and stiff drag, the quality of the numerical solution depends on the ratio between the smoothing length and cell size. For one-dimensional test problems, considered in this paper, we have found, that an optimal smoothness is achieved for $h_{\rm cell}=0.5h$.

\begin{figure*}
\includegraphics[width=0.8 \textwidth]{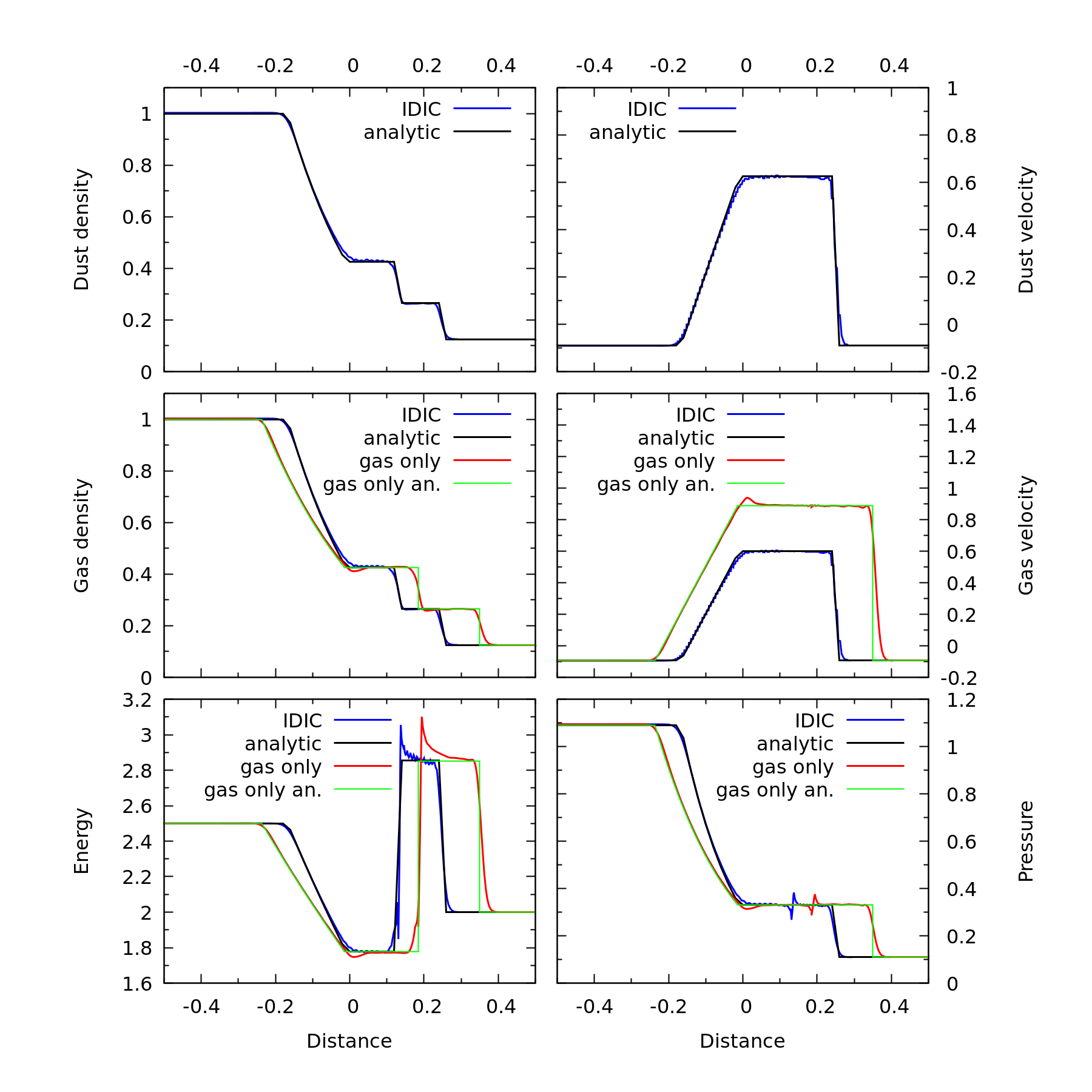} 
\caption{Solution of the DustyShock problem by the SPH-IDIC scheme for the time moment $t=0.2$ with extremely stiff drag $(K=5000)$ and high concentration of dust in the mixture $\varepsilon=1$. Total number of particles, involved in the computation $N_{\rm total}=2 \times 990$, $h=0.01$, $\tau=0.001$. Red and green lines show the solution of the shock tube problem in gas with $N_{\rm gas}=990$ and the same $h$ and $\tau$.} 
\label{fig:d2g1_K5000_cells} 
\end{figure*}

Fig.\ref{fig:d2g1_K5000_cells} shows the numerical and analytical solution for the DustyShock problem at a time moment $t=0.2$ for the gas-dust medium with extremely stiff drag $K=5000$, thus $t_{\rm stop}=0.0002$, and high dust concentration. The numerical solution is obtained from the SPH-IDIC scheme with time step  $\tau \gg t_{\rm stop}$, besides that, $h_{\rm cell}=0.5h$. It is seen, that at $h=0.01$, $N=990$ the numerical solution for the shock and rarefaction waves in the gas medium is close to the analytical one, the dissipation is manifested only weakly (for comparison, see two left panels in Fig.\ref{fig:d2g1_K500_shock}). However, in the vicinity of the contact discontinuity, both internal energy and pressure drops occur. The accuracy of the internal energy reproduction is about 10\%. 

A numerical drop in the internal energy (and hence - a pressure drop as a consequence) on the contact discontinuity is a feature of the Smoothed particles hydrodynamics in combination with the first order approximation scheme by time (see, e.g. \citet{ChaGSPH2003}, Figs. 11-14 there). To minimize the method biases for the discontinuous solutions several modifications of SPH are developed (see e.g. \citet{ChaGSPH2003}). In the paper we just ensured, that the SPH-IDIC method for a gas-dust medium does not increase the error of the solution around the contact discontinuity. To this end we compared the numerical and analytical solution of the shock tube problem in a gas with given above initial data and parameters. The results are visualized on four bottom panels of Fig.\ref{fig:d2g1_K5000_cells}. On the bottom left panel of Fig.\ref{fig:d2g1_K5000_cells} it is seen, that the internal energy drop is the same for gas and gas-dust medium. 

Moreover, the dissipation for the wave fronts in the gas-dust medium with high dust concentration and stiff drag is manifested weaker compared to the gas medium. The dissipation of wave fronts in gas is a consequence of the introduction of the artificial viscosity into the equations of gas motion. However, the artificial viscosity is not included into the equations for dust. As a result of high drag in a medium with high dust concentration, the gas velocity appears to be close to the dust velocity, which decreases the dissipation.

\subsection{Kernel and accuracy comparison}
\label{sec:kernels}

\begin{figure*}
\center
\includegraphics[width=0.9 \textwidth]{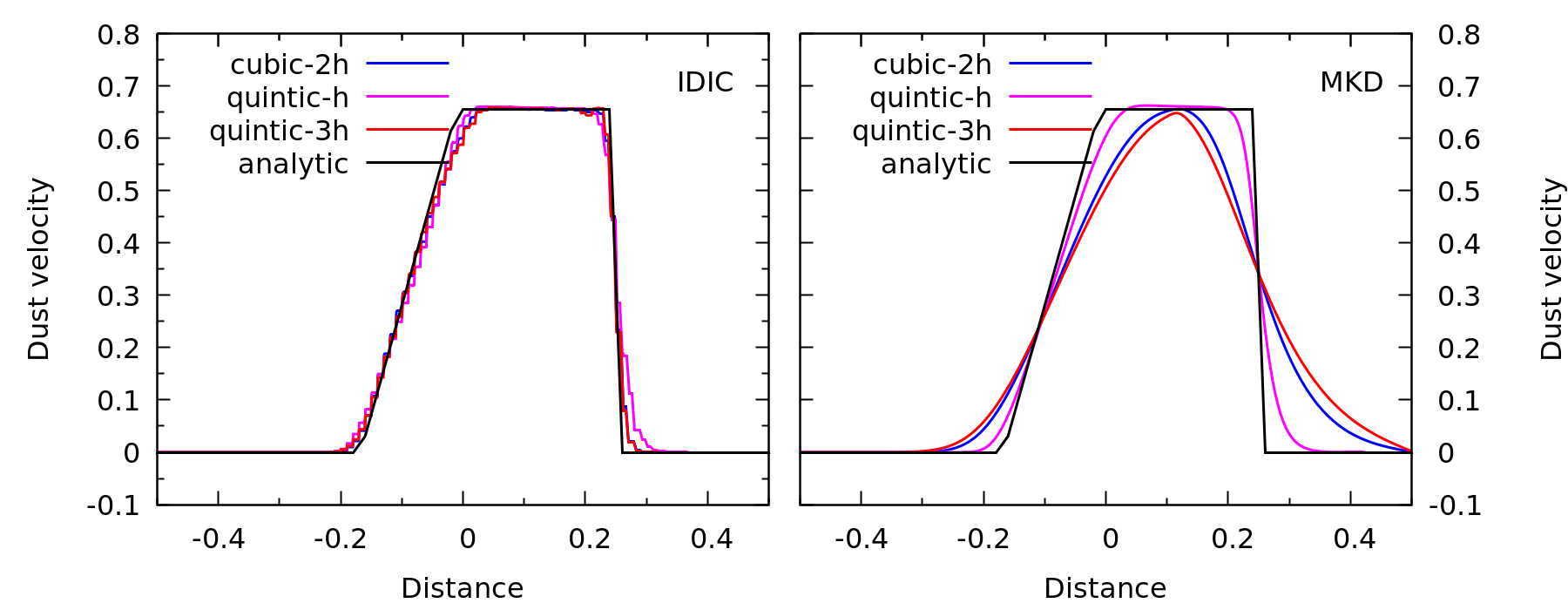}
\caption{Numerical and reference solution of DustyShock problem with stiff drag $K=500$ and high dust concentration $\varepsilon=1$. Standard numerical parameters $N_{\rm total}=2 \times 990$, $h=0.01$, $\tau=0.1$ are used. Dust velocity in time moment $t=0.2$ is plotted. Results with kernels of different support domain radius is given for MKD (right panel) and IDIC scheme (left panel).} 
\label{fig:shockL2} 
\end{figure*}

To study the influence of kernels on the accuracy of results, we compared the solutions of DustyWave and DustyShock problems for the case of stiff drag $K=500$ and high dust concentration $\varepsilon=1$ obtained with cubic spline, quintic-$h$ and quintic-$3h$ kernels. For both problems standard numbers of SPH particles are used, values for $\tau$ and $h$ are given in Tabs. \ref{tab:L2dustyshock},\ref{tab:L2dustywave}. The results are shown on Fig.\ref{fig:shockL2}. Visual examination of right panel of Fig.\ref{fig:shockL2} indicates, that the wider is the kernel support domain, the more dissipation of solution is obtained with MK scheme. This tendency is in agreement with results of e.g. \cite{LaibePrice2011Test}. The left panel of Fig.\ref{fig:shockL2} demonstrates that IDIC scheme significantly decreases the level of dissipation when wide kernels (cubic spline and quintic-3h) are implemented. As it follows form Tab.\ref{tab:L2dustyshock} transition from MKD to IDIC way of drag force computing makes the scheme error 3 times smaller for quintic-3h kernel and only 30 \% smaller for quintic-h kernel. Moreover, the same tendency in more exaggerated form can be seen from Tab.\ref{tab:L2dustywave}: DustyWave solution becomes 30-100 times more accurate when IDIC scheme is implemented for quintic-3h kernel, and doesn't change for quintic-h kernel. 

Moreover, one can see from Tab.\ref{tab:L2dustyshock},\ref{tab:L2dustywave} that synergy of wide kernel with IDIC way of drag term computing provides more accurate results than just implementation of narrow kernel for MK scheme on both test problems. 

\begin{table}
\small
\caption{Scheme error calculated for dust velocity with $L_2$ norm for DUSTYSHOCK problem with stiff drag $K=500$ and high dust concentration $\varepsilon=1$. $M_4$ is the cubic spline kernel, $M_6$ is the quintic kernel. $N_{total}=2\times990$ SPH particles.}
\begin{center}
\begin{tabular}{|c|c|c|c|c|c|}
\hline & & & $M_4$-2h & $M_6$-h & $M_6$-3h \\
\hline \multirow{2}{*}{$h=0.01$} & $\tau=0.0001$ &MK & 0.138 & 0.074 & 0.156 \\ &$\tau=0.001$ &IDIC & 0.051 & 0.063 & 0.049 \\ 
\hline \multirow{2}{*}{$h=0.025$} & $\tau=0.00025$ &MK & 0.235 & 0.143 & 0.270 \\ & $\tau=0.0025$ &IDIC & 0.094 & 0.116 & 0.092 \\ 
\hline
\end{tabular}
\end{center}
\label{tab:L2dustyshock}
\end{table}

\begin{table}
\small
\caption{Scheme error calculated for dust velocity with $L_2$ norm for DUSTYWAVE problem with stiff drag $K=500$ and high dust concentration $\varepsilon=1$. $M_4$ is the cubic spline kernel, $M_6$ is the quintic kernel. $N_{total}=2\times 600$ SPH particles.}
\begin{center}
\begin{tabular}{|c|c|c|c|c|c|}
\hline & & & $M_4$-2h & $M_6$-h & $M_6$-3h \\
\hline \multirow{2}{*}{$h=0.01$} & $\tau=0.001$ &MK & 0.039 & 0.005 & 0.059 \\ & $\tau=0.001$ &IDIC & 0.002 & 0.006 & 0.002 \\ 
\hline \multirow{2}{*}{$h=0.025$} & $\tau=0.001$ &MK &  0.217 & 0.041 & 0.296 \\ & $\tau=0.0025$ &IDIC & 0.004 & 0.040 & 0.003 \\ 
\hline
\end{tabular}
\end{center}
\label{tab:L2dustywave}
\end{table}

\subsection{Discussion of the results} 

\subsubsection{Conservative properties of compared numerical schemes} 

In this Section we consider the principal differences in the properties of the involved schemes, which, to our opinion, result in essentially different level of dissipation at stiff drag. We suppose, that this difference consists in the conservative properties of the used numerical schemes, which is related to the law of momentum conservation as one of the principal conservation laws for physical quantities.

First, we shall show, that for the system (\ref{eq:gas})-(\ref{eq:dust}) the law of the momentum conservation is valid locally for an arbitrary Eulerian volume. Assuming, that 
\begin{equation} 
a_{\rm g}=-\displaystyle\frac{\nabla P}{\rho_{\rm g}}+g, \ \ a_{\rm d}=g, 
\end{equation} 
from (\ref{eq:gas})-(\ref{eq:dust}) we obtain the system for $u$ and $v$ 
\begin{equation} 
\label{eq:system} 
\displaystyle \frac{\partial v}{\partial t} = a_{\rm g}-\varepsilon\displaystyle\frac{v-u}{t_{\rm stop}}, 
\quad\displaystyle \frac{\partial u}{\partial  t} = a_{\rm d} + \displaystyle\frac{v-u}{t_{\rm stop}}. 
\end{equation} 
If $a_{\rm g}=0$, $a_{\rm d}=0$ and $\varepsilon=const$, which corresponds to a limiting value of  $t_{\rm stop}=0$, it follows from (\ref{eq:system}), that 
\begin{equation} 
\label{eq:conserv} 
\displaystyle\frac{\partial }{\partial  t}(v+\varepsilon u)=0. 
\end{equation}

Evidently, if (\ref{eq:conserv}) is valid for an arbitrary Eulerian volume, it is valid for the whole domain. 

In the MK scheme the calculation of the drag force is performed on the base of a pairwise symmetric interaction of gas and dust particles. The formulas are designed to ensure that the momentum is exactly conserved for the whole computational domain in the numerical solution, that is 
\begin{equation} 
\label{eq:voltotalmoment} 
m_{\rm g} \displaystyle \sum_{a} \left[\frac{\mathrm{d} v_a}{\mathrm{d} t}\right]_{\rm drag}+ m_{\rm d} 
\displaystyle \sum_{j} \left[\frac{\mathrm{d} u_j}{\mathrm{d} t}\right]_{\rm drag}=0, 
\end{equation} 
where the summation by $a$ and $j$ is performed for all particles. It is evident, that (\ref{eq:voltotalmoment}) is a discrete analog of (\ref{eq:conserv}), if the whole domain is taken as a volume in (\ref{eq:conserv}). However, this does not imply the fulfillment of the condition (\ref{eq:voltotalmoment}) for an arbitrarily selected volume.

In the SPH-IDIC scheme the drag is calculated using particle allocation into Eulerian volumes or cells. For each of the selected cell the fulfillment of (\ref{eq:voltotalmoment}) with summation over all particles in this cell takes place. Indeed, if we set 
\begin{equation} 
\displaystyle\left[\frac{\mathrm{d} v_i}{\mathrm{d} t}\right]_{\rm drag}=\varepsilon^* \frac{u_*-v_i}{t^*_{\rm stop}}, \quad \displaystyle\left[\frac{\mathrm{d} u_j}{\mathrm{d} t}\right]_{\rm drag}=\frac{u_j-v_*}{t^*_{\rm stop}}, 
\end{equation} 
then 
\begin{equation} 
\label{eq:totalV} 
\displaystyle \sum_{i=1}^N \left[\frac{\mathrm{d} v_i}{\mathrm{d} t}\right]_{\rm drag}=\frac{\varepsilon^*}{t^*_{\rm stop}}\left(Nu_*-\sum^N_{i=1}v_i \right)=\frac{\varepsilon^* N}{t^*_{\rm stop}}\left(u_*-v_* \right), 
\end{equation} 
\begin{equation} 
\label{eq:totalU} 
\displaystyle \sum_{j=1}^L \left[\frac{\mathrm{d} u_j}{\mathrm{d} t}\right]_{\rm drag}= \frac{1}{t^*_{\rm stop}} 
\left( \sum_{j=1}^L u_j - L v_*\right)=\frac{L}{t^*_{\rm stop}}(u_*-v_*). 
\end{equation} 
From (\ref{eq:aveEpsilon}),(\ref{eq:totalV}) and (\ref{eq:totalU}) the validity of (\ref{eq:voltotalmoment}) for the considered cells follows directly.

The obtained results point at the fact, that a scheme, where the momentum conservation law is ensured locally, not only globally, is free from the over-dissipation of the numerical solutions in the tests considered here. 

\subsubsection{Semi-implicit approximation of the drag force and requirements to the time step} 

On the base of MK, IPSPH and DIC approaches it is possible to approximate the drag terms in (\ref{eq:gas})-(\ref{eq:dust}) in both explicit and semi-implicit ways. Under semi-implicit approximation, the drag coefficient and gas and dust densities are taken from the previous time step, and velocities are taken from the next time step. The semi-implicit approximation allows to remove the strict condition (\ref{eq:timeresSPH}) to be imposed on the step $\tau$ in explicit schemes.

For the MK methods the implementation algorithms for semi-implicit computing of the drag force are described in \citet{MonaghanKocharyan1995,FranceDustCode,LaibePriceAstroDrag}. Neither of them is a direct scheme, that is they require converging iterations, and the convergence rate of a particular method depends on the parameters of the drag force. The IPSPH approach ensures the implementation of an implicit scheme with as a direct algorithm with fixed number of arithmetic operations (see \citet{BateDust2014} for details). In \ref{sec:implicitdrag} we have shown, that the DIC approach allows to obtain a direct and fast implementation for the semi-implicit approximation of drag force, if the first order method with respect to time is used, and the system is considered to be in the Epstein mode.

A summary of computational approaches for the drag force is given in Table \ref{tab:resume}. 

\begin{table*} 
\centering
\begin{minipage}{160mm} 
\caption{Characteristics of drag force calculation methods} 
 \label{tab:resume} 
 \begin{tabular}{@{}llll@{}} 
 \hline
Computation method     & Ensures the fulfillment   &    Preserves Lagrangian & Allows for direct \\ 
for the drag force & of the local              &    nature of SPH        & implementation of \\ 
                       & momentum conservation law &                         & semi-implicit drag \\ 
  \hline
    MK & - & + & - \\ 
       &   &   &   \\ 
  \hline
    IPSPH & - & + & + \\ 
          &   &   &    \\ 
  \hline
    DIC & + & - & + \\ 
        &   &   &   \\ 
\end{tabular} 
\end{minipage} 
\end{table*}

\section{Conclusion} 
\label{sec:resume} 

The article considers the novel Lagrange-Eulerian SPH-IDIC method for the calculation of the dynamics of gas and dust mixture on the base of two-fluid smoothed particle hydrodynamics (TFSPH). Comparing to previously developed TFSPH methods SPH-IDIC scheme doesn't require spatial resolution condition $h < c_s t_{\rm {stop}}$ for accurate computing of mixtures with high dust concentration and small grains, which is essential for protoplanetary disc simulations. Below we outlined the idea of SPH-IDIC and our main results.

Within TFSPH, gas and dust phases are simulated with two separate sets of particles. We considered the dust to be monodisperse, that is containing only grains with one typical size. In SPH-IDIC, the pressure force is calculated via a standard SPH approximation. To calculate the drag force, the particles are allocated into nonoverlapping densely spaced volume cells. For each cell, averaged parameters of the gas-dust mixture are calculated. Drag forces, acting on every gas or dust particle, is calculated with use of the individual velocity of a particle and averaged by volume cell characteristics of the gas-dust mixture. The method is designed so, that for the case of linear dependence of the drag force on the relative velocity (i.e. Epstein drag mode) following requirements are met: 

(1) it is ensured, that for every local volume the momentum, lost by the gas due to drag force, will be equal to the momentum, gained by the dust subsystem due to drag force, 

(2) implicit drag is implemented in a computationally fast way. 

We have compared the suggested SPH-IDIC method with an explicit method of \cite{MonaghanKocharyan1995}, which is classical for TFSPH approaches. We have shown, that both methods yield similar results for media with high dust concentration and weak drag (large grains). We have also shown, that for mixtures with high dust concentration and stiff drag (small grains), resolution condition $h<c_{\rm s} t_{\rm stop}$ is crucial to avoid numerical overdissipation for \cite{MonaghanKocharyan1995} scheme but is not required for SPH-IDIC scheme. Moreover, SPH-IDIC scheme decreases the level of dissipation due to artificial viscosity in shock simulations.  

Therefore, the SPH-IDIC method can be considered as a promising tool for the simulation of the dynamics of gas-dust mixture with arbitrary dust concentration and arbitrary sizes of dust grains. The following steps are considered as prospective directions for future research efforts: 

(1) determining the accuracy of numerical solutions depending on the partitioning of the domain into cells, and number of gas and dust particles in a cell. 

(2) testing the method on two-dimensional and three-dimensional problems, determining the most efficient algorithm for partitioning of the computational domain into cells, 

(3) studying the ways of extension of the approach to polydisperse dust media.

\section*{Acknowledgements}

This work was supported by the Boreskov Institute of Catalysis. The authors are grateful to anonymous referee for constructive and useful comments.
\appendix

\section{Implementation of the SPH-IDIC scheme with first order approximation of time} 
\label{sec:implicitdrag} 

Let us denote

\begin{equation*}
\Psi^{n}_a = - \displaystyle \sum_b m_b\left(\frac{P^{n}_b}{(\rho^n_{b,\rm{g}})^2} + \frac{P^{n}_a}{(\rho^n_{a,\rm g})^2} + \Pi_{ab} \right) \bigtriangledown_a W_{ab}.
\end{equation*}

Performing the summation in the equation (\ref{eq:DragInCellv}) by $a$ and dividing the result by the number of gas particles $N$ in a cell, we obtain the equation for $v_*$: 

\begin{equation} 
\label{eq:v_asterisk} 
\frac{v^{n+1}_*-v^{n}_*}{\tau} = \displaystyle \Psi^n_* - \varepsilon^n_* \frac{v^{n+1}_* - u^{n+1}_*}{t^{n, *}_{\rm stop}}, 
\end{equation} 
where $\Psi^n_* = \displaystyle \frac{1}{N} \sum^N_{i=1} \Psi^n_a$.

Similarly for $u_*$: 
\begin{equation} 
\label{eq:u_asterisk} 
\frac{u^{n+1}_*-u^{n}_*}{\tau} = \displaystyle \frac{v^{n+1}_* - u^{n+1}_*}{t^{n, *}_{\rm stop}}. 
\end{equation} 
\\It is convenient to express the solution for (\ref{eq:v_asterisk})-(\ref{eq:u_asterisk}) using the substitution 
\begin{align} 
\label{xy_variables} 
&x^{n+1} = v^{n+1}_* - u^{n+1}_*, \ \, x^n = v^n_* - u^n_*, \nonumber \\ & \displaystyle y^{n+1} = v^{n+1}_* + \varepsilon^n_* u^{n+1}_*,  \ \ \displaystyle y^{n} = v^{n}_* + \varepsilon^n_* u^{n}_*, 
\end{align} 
which results in the following: 
\begin{equation} 
\label{eq:x_asterisk} 
\frac{x^{n+1}-x^n}{\tau} = \Psi^n_* - x^{n+1}(\frac{\varepsilon^n_* + 1}{t^{n, *}_{\rm stop}}), \quad 
\frac{y^{n+1}-y^n}{\tau} = \Psi^n_*, 
\end{equation} 
\begin{equation} 
\label{eq:x_asterisk_final} 
\left(\frac{1}{\tau}+\frac{\varepsilon^n_* + 1}{t^{n, *}_{\rm stop}}\right)x^{n+1} =\frac{x^n}{\tau}+ \Psi^n_* , \quad 
y^{n+1}=y^n +\tau \Psi^n_*, 
\end{equation} 
then 
\begin{equation} 
v^{n+1}_* = \displaystyle \frac{y^{n+1} + \varepsilon^n_* x^{n+1}}{1 + \varepsilon^n_*}, \quad u^{n+1}_* = \displaystyle \frac{y^{n+1} - x^{n+1}}{1 + \varepsilon^n_*}. 
\end{equation} 
Determine from (\ref{eq:DragInCellv})-(\ref{eq:DragInCellu}) the values $v_a^{n+1}$, $u_j^{n+1}$: 
\begin{align} 
\left(\frac{1}{\tau}+\frac{\varepsilon^n_*}{t^{n, *}_{\rm stop}}\right)v^{n+1}_a =\frac{v^n_a}{\tau}+ \frac{\varepsilon^n_*}{t^{n, *}_{\rm stop}} u^{n+1}_* + \Psi^n_* +g_a, \nonumber \\ \left(\frac{1}{\tau}+\frac{1}{t^{n, *}_{\rm stop}}\right)u^{n+1}_j =\frac{u^n_j}{\tau}+ \frac{1}{t^{n, *}_{\rm stop}} v^{n+1}_* + g_j. 
\end{align}

\section*{References}
\bibliography{Stoyanovskaya}

\end{document}